\documentclass[aps,prc,twocolumn,groupedaddress,floatfix]{revtex4}
\usepackage{graphicx}
\usepackage{dcolumn}

\begin{document}


\title{Effective lagrangian approach to structure of selected nuclei far from stability}


\author{M. A. Huertas}
\email[]{huertas@camelot.physics.wm.edu}
\affiliation{Department of PhysicsCollege of William and Mary, Williamsburg, VA 23187}


\date{\today}

\begin{abstract}
The convergence of the effective field theory (EFT) approach of Furnstahl, Serot and Tang to the nuclear many-body problem is studied by applying it to selected doubly-magic nuclei far from stability. An independently developed code, which can incorporate various levels of approximation of the chiral effective lagrangian, is used to solve the self-consistent relativistic Hartree equations. Results are obtained for ground-state properties such as binding energies, single-particle level structure, and densities of the selected spherical, doubly-magic nuclei $^{132,100}\textrm{Sn}$ and $^{48,78}\textrm{Ni}$. Calculated spectra of neighboring nuclei differing by one particle or one hole show agreement with the most recent experimental data. Predictions for nucleon densities are presented.
\end{abstract}


\maketitle

\section{Introduction}
The study of nuclei far from the $\beta$-stability line has drawn the attention of nuclear physicists. The implications of the structure and reaction cross-sections from these nuclei play an important role in the description of nucleosynthesis processes such as the r-process, which takes place in the region of neutron rich nuclei. In addition to this, these nuclei are the perfect place to test the validity of some successful methods and theories, which have rendered satisfactory results along the stability line for medium-heavy nuclei. The development of new experimental facilities and the implementation of new techniques have already given experimental data for some of these, like the doubly-magic nuclei $^{132}_{\ 50}\textrm{Sn}_{82}$ and $^{100}_{\ 50}\textrm{Sn}_{50}$. Some of the new approaches being tested are those of relativistic mean-field models. In this paper we are going to deal with a relativistic mean-field theory derived from the chiral effective lagrangian of Furnstahl, Serot and Tang~\cite{CEL}. This lagrangian, which incorporates the symmetries of QCD, particularly a non-linear realization of chiral symmetry, is built as an effective field theory and contains the lowest lying hadronic degrees of freedom as the main ingredients. Though it reproduces the ground-state properties of stable doubly-magic nuclei, its predictability far from the $\beta$-stability region has not been tested, especially the convergence of the calculations as the number of effective terms in the lagrangian increases.

Relativistic mean-field models can account for a large number of bulk properties of nuclei and there are different methods that have produced acceptable results. Recently, Toki, \textit{et al.}~\cite{Toki} used a relativistic mean-field approach to calculate properties of nuclei from the proton to the neutron drip lines, extending the calculations to super-heavy nuclei. The good agreement with known experimental data shows how relativistic mean-field models can account for a variety of phenomena in the nuclear many-body system. Toki and his co-workers, using a \textit{mean effective lagrangian}, have an extensive program underway on applications of this approach to astrophysics, in particular to element formation in supernovas~\cite{Toki-2}. 

The idea of using effective field theories to describe the nuclear problem is an increasing field of study, an account of recent results can be found in~\cite{Furnstahl1}. The relativistic mean-field approach to the nuclear many-body problem has been successful in describing medium-size nuclei, and its success can be understood under the perspective of effective field theory (EFT) and density functional theory (DFT)~\cite{Furnstahl, Ser-Wal}. The success of previous relativistic mean-field models, like the original Walecka model~\cite{Wal-book, Ser-Wal} and its extensions to include nonlinear couplings of the $\sigma$ field, is now understood by applying the ideas of naturalness from EFT~\cite{Furnstahl2}. A comparison of various mean-field models and an account of their success was presented by Furnstahl, et al~\cite{Furnstahl3}. In that paper it was shown, using the EFT approach and NDA, why various mean-field approaches have been successful and that the use of different degrees of freedom to describe nuclear properties corresponds to different organizational principles. Thus a description of the nuclear many-body system based on nucleon and meson fields, as in this approach, or an expansion in nucleon densities, as in the Skyrme model, are equivalent. The principles of EFT rely on the fact that there exist natural scales and in order to understand a particular phenomenon it is enough to use probes that can resolve the dynamics at that scale. These probes are the relevant degrees of freedom needed to describe the system. The most general lagrangian density is built from these interacting fields paying attention to symmetries imposed by an underlying theory (QCD in this case) and arranged by some organizing principle, like the naive dimensional analysis (NDA) from Georgi and Manohar~\cite{NDA-1,NDA-2}. The unknown dynamics, of heavier degrees of freedom, is integrated out and appears as coupling constants in the theory. These are determined by fitting the theory to known experimental data. The number of terms in this EFT lagrangian is infinite and,  in order to give predictive power to the model, it is necessary to have an organizing principle that groups together terms that contribute at the same level. In this way it is possible to truncate the expansion at some arbitrary order and be certain that all non-redundant terms, of the same order, have already been taken into account. 

The inclusion of new interaction terms in the lagrangian aims to give a better description of the energy functional, thus going beyond the Hartree approximation and incorporating exchange-correlation effects. The new terms are non-renormalizable, a distinction with standard RMF models, which truncate the effective lagrangian at this level without any underlying physical justification. A chiral effective lagrangian has been derived by Furnstahl, Serot, and Tang~\cite{CEL}, which incorporates a non-linear realization of chirial symmetry and uses the lowest lying hadronic levels: i) pions, whose field appears as the phase of a chiral rotation of the identity matrix, ii) nucleons, described by isospinor field N,  iii) an isoscalar-vector field, whose mass is taken to be equal to the $\omega$-meson mass and iv) an isoscalar-scalar field which simulates correlated two-pion exchange with an effective mass around 500 MeV, and v) an isovector field, the $\rho$-meson, to take into account the symmetry energy in nuclei. By applying the relativistic mean-field approximation, a set of equations of motion can be derived from the effective lagrangian in which the meson fields act as classical fields. 

The connection to DFT is made by considering the expansion of the relativistic mean-field lagrangian as an expansion of the energy functional of the nuclear many-body problem in terms of nucleon densities and auxiliary classical meson fields. According to the principles of DFT~\cite{Kohn,DFT-QHD-1}, knowledge of the exact energy functional can be used, by optimizing it with respect to the fermion densities and auxiliary potentials, to calculate the exact ground-state scalar and vector densities, energy, and chemical potential for the fully interacting many-fermion system~\cite{Many-body}. These optimal observables are used here to compare the experimental values with the calculated ones. Other observables are not expected to be well reproduced, but a comparison with some of them is included in this paper in order to understand the extent to which the theory can be applied, as well as to find new directions in which more work needs to be done.
 In this way the EFT principles combined with DFT methods give an approach to the nuclear many-body problem that uses the simplicity of solving the self-consistent Hartree equations in which new contributions coming from the interacting fields can be incorporated in a systematic and controlled way. The expansion parameters that make the truncation of the effective lagrangian possible in the nuclear structure case are essentially the ratios of mean isoscalar, scalar and vector fields to the nucleon mass $(\approx 1/3)$ and the ratio of the Fermi momentum to the nucleon mass $(\approx 1/4)$.

In this paper we study the \textit{convergence} of the chiral effective lagrangian (CEL) approach to the nuclear many-body problem when extrapolated to selected doubly-magic nuclei far from the $\beta$-stability region. The study focuses at how well additional levels of approximation of the energy functional reproduce the above ground-state properties of the selected, spherical, doubly-magic nuclei. In addition to this, we also examine the effect of fitting a larger set of observables, in the mean, along the stability line. We center our attention on the four doubly-magic nuclei $^{132}_{\ 50}\textrm{Sn}_{82}$, $^{100}_{\ 50}\textrm{Sn}_{50}$, $^{48}_{28}\textrm{Ni}_{20}$ and $^{78}_{28}\textrm{Ni}_{50}$ chosen for their relevance to the nucleosynthesis processes. Here we compare the predictions of the CEL model at various levels of approximation with the most recent experimental results for these nuclei. We compare results of nuclear binding energies for these nuclei and also extend this analysis to neighboring nuclei that differ by one particle or one hole from the doubly-magic nuclei thus comparing the values of the chemical potential. We find in both cases remarkable agreement with recent experimental data. In addition to this we look at the convergence for single-particle levels, both for neutrons and protons. We examine how sensitive these calculated spectra are to the various levels of approximation in the CEL, as new interaction terms come into play.  Nuclear densities are also shown for the ``proton rich'' nucleus $^{100}_{\ 50}\textrm{Sn}_{50}$ and the neutron rich one $^{132}_{\ 50}\textrm{Sn}_{82}$. These densities are in very good agreement with other calculations~\cite{Lalazissis} using the relativistic Hartree Bogoliubov equations. According to the EFT approach used here, the systematic inclusion of interaction terms will approach the \textit{true energy functional}, going beyond the Hartree level and including exchange-correlation terms. Thus, here we include additional pairing interactions  only to the extent they are systematically included through the additional terms in the CEL and the fitting of the constants. The goal of the present work is to study how well the current level of approximation in the CEL reproduces the ground-state properties of the selected nuclei without explicit use of the pairing interactions.

The calculations are done using an independently developed code, which solves the self-consistent Hartree equations at any level of approximation. The code has been tested successfully by reproducing previous calculations. The coupling constants, for each level of approximation, were obtained by Furnstahl, \textit{et al.}~\cite{CEL} by fitting in the mean to properties of stable doubly-magic nuclei leading to different parameter sets according to the level of approximation~\cite{CEL, Ser-Wal}. We use here these results and use the same name for the parameter sets, i.e. W1, Q1, Q2, G1 and G2 in increasing level of approximation, respectively. In addition to these, we include two additional sets: L2 and NLC. The parameter set L2 corresponds to the same level of approximation as the set W1, it contains no non-linear terms in the meson fields. The other parameter set, NLC, has the same level of approximation as Q1. Both L2 and NLC were fit to describe different nuclear properties than those corresponding to the sets W1, Q1, Q2, G1 and G2 \protect\footnote{The NLC set describes the deformation of selected medium-weight nuclei~\cite{deformed}}. They also differ from these in the number of nuclei used to fit the parameters. The values of the constants used in this paper are given in Table~\ref{tab:parms}, and the fitting procedure is discussed in more detail later in this paper.

Other approaches have been applied to studies of nuclei far from stability with various degrees of success. Neutron and proton skin formation, the reduction of the spin-orbit interaction and the realization of pseudo-spin symmetry are some of the phenomena studied recently by various authors~\cite{Lenske,Meng,skins}. Shell-model calculations using the CD-Bonn effective potential modified by the G-matrix and renormalized using the Q-box method have been used to study the importance of the $^{1}\textrm{S}_{0}$ and $^{3}\textrm{P}_{2}$ partial waves in reproducing the constancy of the $2^{+}_{1}-0^{+}_{1}$ level spacing which is apparent in Sb nuclei~\cite{England}. Some of the methods that have been tested on these nuclei with various degrees of success are the density dependent relativistic Hartree (DDRH)~\cite{Hofmann,Lenske}, relativistic continuum Hartree Bogoliubov (RCHB)~\cite{Meng}, relativistic Hartree Bogoliubov (RHB)~\cite{Lalazissis}, Hartree-Fock-Bogoliubov~\cite{Grasso} and relativistic Dirac-Hartree-Bogoliubov (DHB)~\cite{Carlson}. In these methods pairing correlations have been taken into account using a phenomenological force like the Gogny force~\cite{Meng,Lalazissis} or a density-dependent zero-range interaction~\cite{Hofmann}. Other authors~\cite{Estal,Hofmann} used the BCS approach with a constant pairing strength to take into account the effects of the continuum. Perhaps the most self-consistent approach has been the use of the RHB, in which ph-interactions and pairing correlations have been taken into account in a unified way by solving self-consistently the RHB-equations with the Klein-Gordon equations that describe the meson fields~\cite{Lalazissis,Carlson}. Other authors have derived new parameter sets of known potentials, e.g. Nilsson~\cite{Zhang}, Groningen~\cite{Hofmann}, to fit the new experimental data of binding energies, single-particle levels, and nuclear radii for these nuclei far from stability. Relativistic mean field models (RMF), which have been successful in describing bulk properties of stable nuclei, are being tested in these new regimes. Some of the models use the $\sigma-\omega$ lagrangian with linear and non-linear $\sigma$ interactions~\cite{Lalazissis, Zhang}. Applications of DFT to the nuclear many-body problem have been carried out using the QHD-I and QHD-II lagrangians~\cite{DFT-QHD-1, DFT-QHD-2, DFT-QHD-3}. Here the exchange-correlations are taken directly from the applications of many-body techniques in infinite nuclear matter, but extensions to nuclei far from stability have not yet been performed in this approach. Direct application of the QHD-II lagrangian to the doubly-magic Sn isotopes to calclulate single-particle level spectrum and nucleon densities can be found in~\cite{Bachman}.

The combination of the fundamental idea of DFT and the realization that an approximation to the exact energy functional can be derived using EFT methods is a new approach to the nuclear many-body problem. As such, it is still in development. To the extent that a sufficient number of terms are retained in the effective lagrangian to provide the correct energy functional, one has actually developed the effective field theory for QCD at the nuclear density domain. Extensions to other density regimes still has to be explored. This paper explores how additional terms in the effective lagrangian contribute to the results for ground-state properties of selected nuclei far from stability. The EFT methods in this chiral effective lagrangian approach have already shown that the coupling constants are all natural~\cite{CEL}, therefore the relevant degrees of freedom have been correctly taken into account and the NDA works. Knowing how well this EFT approach behaves at these extreme cases of nuclei far from stability will give a firmer ground for the theory. Predictions for the properties of these nuclei is also, of course, of direct relevance to experiments, both ongoing and planned~\cite{GSI}, and to element formation in supernovas.

In this paper we see that the increasing level of approximation approaches the total binding energy of the selected nuclei. There is also very good agreement with the binding energies of neighboring nuclei, which differ by one particle or one hole from the doubly-magic nucleus, i.e. the chemical potential. The spins and parities of these nuclei, to the extent to which they have been measured, are also correctly given. In addition, we show the calculated single-particle excitation spectra and examine the effects of additional levels of approximation. Finally, we present the nucleon densities for neutron rich and neutron deficient nuclei predicted with this approach.

The following two sections explain the effective lagrangian used in this paper and describe the calculations, and in the last two sections results are presented and conclusions are drawn.

\section{Calculations}

The chiral effective lagrangian used here was constructed using the relevant lowest lying hadronic degrees of freedom~\cite{CEL}. The pion field is introduced as the phase of a chiral rotation of the identity matrix in isospin space. A field $\xi(x)$ is defined in the following way:
\[
\xi(x) = exp(i\pi(x)/f_{\pi}).
\]
where the pion field $\pi(x)$ is defined by $\pi(x)=\frac{1}{2}\mathbf{\vec \tau \cdot \vec \pi}$

Nucleons are included as isospinor fields, $N(x)$, in which the upper component corresponds to the proton and the lower component the neutron.
\[
N = \left( \begin{array}{c} p(x) \\ n(x) \end{array} \right)
\]
To include a contribution to the symmetry energy of nuclear matter, a $\rho_{\mu}$-field is added to the lagrangian. 

A non-linear realization of chiral symmetry is considered following the work of Callan, Coleman, Wess, and Zumino~\cite{CCWZ}. A description of the incorporation of chiral symmetry in the nuclear many-body lagrangian can be found in~\cite{Wal-book,Ser-Wal}. 
In this formalism, the non-linear realization of the chiral symmetry is defined by a global transformation, L and R, of the subgroups $SU(2)_{L}$ and $SU(2)_{R}$ such that
\begin{equation}
L\otimes R : \quad (\xi,\rho_{\mu},N)\to(\xi',\rho_{\mu}',N')
\end{equation}
where
\begin{eqnarray}
\xi'(x) &=& L \xi(x) h^{\dagger}(x) = h(x) \xi(x) R^{\dagger}
               \ , \label{eq:Xitrans} \\[4pt]
\rho'_\mu(x) &=& h(x) \rho_\mu (x) h^{\dagger}(x)
               \ , \label{eq:Rhotrans} \\[4pt]
 N'(x) &=& h(x)N(x)  \ .       \label{eq:Ntrans}
\end{eqnarray}
here $\rho_{\mu}(x)=\frac{1}{2}\mathbf{\vec \tau \cdot \vec \rho_{\mu}}$.

In addition to the above fields, an electromagnetic field $A_{\mu}$ is included and the electromagnetic structure of the nucleons has been taken into account. Their anomalous magnetic moments, labeled as $\lambda_{p}$ and $\lambda_{n}$ for proton and neutron respectively, then enter. An isoscalar-vector field $V_{\mu}$ is needed to correctly represent the bulk properties of nuclei. This field is treated as a chiral singlet and its mass is taken to be equal to the  $\omega$-meson mass. Finally, an isoscalar-scalar field $\phi$, again a chiral singlet, is included to incorporate the observed mid-range N-N attraction. Its mass is determined by fitting the calculations to experimental results.

The various interaction terms present in the effective lagrangian have been arranged following the NDA from Georgi and Manohar~\cite{NDA-1,NDA-2}. Details concerning the construction of the chiral effective lagrangian can be found in~\cite{CEL} where the full effective lagrangian is developed.

In order to solve the equations of motion derived from this lagrangian we apply the relativistic mean-field approximation in which the meson fields become expectation values and thus are treated as classical fields. In this approximation the pseudo-scalar pion field does not contribute because it has no expectation value. The interplay between these fields, particularly the scalar and vector fields, accounts for the bulk properties of nuclei as is well described in~\cite{Wal-book,Ser-Wal}. For spherically symmetric systems, as is the case for double-magic nuclei, the spatial components of the vector field vanish. The rho isovector field develops only the non-charged component, $\rho_{0}$, since conservation of charge is imposed. The resulting equations of motion consist of a Dirac equation for the nucleons and a system of non-linear, coupled differential equations for the meson fields.

The Dirac hamiltonian for the nucleon fields takes the form:
{\setlength\arraycolsep{2pt}
\begin{eqnarray}
\label{eq:Dirac-equation}
h(x) & = &  -i\mathbf{\vec{\alpha} \cdot \vec{\nabla}} + W(x) + \frac{1}{2}\tau_{3}R(x) + \beta(M-\Phi(x)) \nonumber \\
& & {}+\frac{1}{2}(1+\tau_{3})A(x){} \nonumber \\
& & {}-\frac{i}{2M} \beta \vec{\alpha} \cdot (f_{\rho} \frac{1}{2} \tau_{3} \vec{\nabla} R + f_{v} \vec{\nabla} W) \nonumber \\
& & {}+\frac{1}{2M^{2}} (\beta_{s} + \beta_{v} \tau_{3})\nabla^{2} A{}\nonumber \\
& & {} - \frac{i}{2M} \lambda \beta \mathbf{\vec{\alpha} \cdot \vec{\nabla}} A       {}
\end{eqnarray}}
Here $\lambda=\frac{1}{2}\lambda_{p}(1+\tau_{3})+\frac{1}{2}\lambda_{n}(1-\tau_{3})$ and the numerical values used for the anomalous magnetic moments are $\lambda_{p}=1.793$, $\lambda_{n}=-1.913$.

The mean meson fields are denoted by $W=g_{v}V_{0}$, $\Phi=g_{s}\phi_{0}$, $R=g_{\rho}b_{0}$, and $A=eA_{0}$ respectively. The quantities ${f_{\rho},f_{v},\beta_{s},\beta_{v}}$ are parameters fit to experiment (see below).
The meson field equations become the following:
{\setlength\arraycolsep{2pt}
\begin{eqnarray}
\label{eq:Phi-equation}
- \nabla^{2}\Phi + m_{s}^{2}\Phi & = & g_{s}^{2}\rho_{s}(x) - \frac{m_{s}^{2}}{M}\Phi^{2} (\frac{\kappa_{3}}{2} + \frac{\kappa_{4}}{3!}\frac{\Phi}{M}){} \nonumber \\
	& & {}+\frac{g_{s}^{2}}{2M}(\eta_{1} + \eta_{2} \frac{\Phi}{M}) \frac{m_{v}^{2}}{g_{v}^{2}}W^{2} {} \nonumber \\
	& & {}+\frac{g_{s}^{2} \eta_{\rho}}{2M} \frac{m_{\rho}^{2}}{g_{\rho}^{2}} R^{2} + \frac{\alpha_{1}}{2M} [(\vec{\nabla} \Phi)^{2} + 2 \Phi \nabla^{2} \Phi] {} \nonumber \\
& & {}+ \frac{\alpha_{2} g_{s}^{2}}{2Mg_{v}^{2}} (\vec{\nabla} W)^{2} {}
\end{eqnarray}}
Here $\rho_{s}$ is the baryon Lorenz scalar density and $g_{s}$, $m_{s}$, $\kappa_{3}$, $\kappa_{4}$, $\eta_{1}$, $\eta_{2}$, $g_{v}$, $g_{\rho}$, $\alpha_{1}$, $\alpha_{2}$ are again parameters fit to experiment. Furthermore,
{\setlength\arraycolsep{2pt}
\begin{eqnarray}
\label{eq:W-equation}
- \nabla^{2}W + m_{v}^{2}W & = & g^{2}_{v}[\rho_{B}(x)+\frac{f_{V}}{2M}\mathbf{\vec{\nabla}\cdot} (\rho^{T}_{B}(x)\mathbf{\hat{r}})] \nonumber \\
& & {}-(\eta_{1}+\frac{\eta_{2}}{2}\frac{\Phi}{M}) \frac{\Phi}{M}m^{2}_{v} W {}\nonumber \\
 & &{}-\frac{1}{3!}\zeta_{0}W^{3} + \frac{\alpha_{2}}{M}(\vec{\nabla} \Phi \cdot \vec{\nabla} W + \Phi \nabla^{2} W) {}\nonumber \\
 & &{}- \frac{e^{2}g_{v}}{3g_{\gamma}}\rho_{\mathrm{chg}}(x) {}
\end{eqnarray}}
Here $\rho_{B}$ is the baryon density, $\rho^{T}_{B}$ is the baryon tensor density, $\rho_{\mathrm{chg}}$ is the charge density and $f_{v}$, $\zeta_{0}$ are parameters. In addition,
{\setlength\arraycolsep{2pt}
\begin{eqnarray}
\label{eq:R-equation}
- \nabla^{2} R + m^{2}_{\rho} R & = &\frac{1}{2} g^{2}_{\rho}[\rho_{3}(x)+\frac{f_{\rho}}{2M}\mathbf{\vec{\nabla}\cdot} (\rho^{T}_{3}(x)\mathbf{\hat{r}})] \nonumber \\
& & {}-\eta_{\rho}\frac{\Phi}{M}m^{2}_{\rho}R-\frac{e^{2}g_{\rho}}{g_{\gamma}}\rho_{\mathrm{chg}}(x) {}
\end{eqnarray}}
Here $\rho_{3}$ and $\rho^{T}_{3}$ are the isovector densities, $f_{\rho}$, $\eta_{\rho}$, $g_{\rho}$ are parameters and $g_{\gamma}$=5.01, is the coupling of the photon to the $\omega$-meson. Finally,
\begin{equation}
\label{eq:A-equation}
- \nabla^{2}A = e^{2}\rho_{\mathrm{chg}}(x) 
\end{equation}
where $\rho_{\mathrm{chg}}$ is the charge density.
To solve the Dirac equation and obtain the single-particle energy levels one has
\begin{equation}
\label{eq:Single-levels}
h\psi_{\alpha}(x) = E_{\alpha} \psi_{\alpha}(x)
\end{equation}

The Dirac solutions are expressed as, 
\begin{equation}
\label{eq:Nucleon-field}
\psi_{\alpha}(x) = 
\left( 
\begin{array}{c}
\frac{i}{r} G_{a}(r) \Phi_{\kappa m} \\
-\frac{1}{r} F_{a}(r) \Phi_{-\kappa m}
\end{array}
\right) \zeta_{t}
\end{equation}
where the $\Phi_{\kappa m}$ is a spin spherical harmonic and t is equal to 1/2 for protons and -1/2 for neutrons. Inserting this into Eq.(\ref{eq:Single-levels}) and using Eq.(\ref{eq:Dirac-equation}) a set of two coupled first-order differential equations for the G and F functions is obtained:

\begin{equation}
\label{eq:G-equation}
\left(\frac{d}{dr} + \frac{\kappa}{r}\right)G_{a}(r) - [E_{a} - U_{1}(r) + U_{2}(r)]F_{a}(r) - U_{3}G_{a}(r) = 0
\end{equation}
\begin{equation}
\label{eq:F-equation}
\left(\frac{d}{dr} - \frac{\kappa}{r}\right)F_{a}(r) + [E_{a} - U_{1}(r) - U_{2}(r)]G_{a}(r) + U_{3}F_{a}(r) = 0
\end{equation}
where the single-particle potentials are given by
\begin{eqnarray}
U_{1}(r) & \equiv & W(r)+t_{a}R(r)+(t_{a}+\frac{1}{2})A(r){}\nonumber \\
 & &{}+\frac{1}{2M^{2}}(\beta_{s}+2t_{a}\beta_{V})\nabla^{2}A(r) {} \\
U_{2}(r) & \equiv & M- \Phi(r) {}\\
U_{3}(r) & \equiv & \frac{1}{2M} \{ f_{V}W'(r)+t_{a}f_{\rho}R'(r){} \nonumber \\
& &{}+A'(r)[(\lambda_{p}+\lambda_{n})/2+t_{a}(\lambda_{p}-\lambda_{n})] \} {} 
\end{eqnarray}
Here the prime indicates a radial derivative; \textit{e.g.} $W^{`}(r)=dW(r)/dr$.

Once the single-particle wave functions are calculated, the various densities that appear on the r.h.s. of the meson equations can be obtained. They are defined as follows:
\begin{eqnarray}
\rho_{s}(x) & = & \sum_{\alpha}^{occ}\frac{2j_{a}+1}{4\pi r^{2}} \left( G^{2}_{a}(r)-F^{2}_{a}(r) \right) {} \\
\rho_{B}(x) & = & \sum_{\alpha}^{occ}\frac{2j_{a}+1}{4\pi r^{2}} \left( G^{2}_{a}(r)+F^{2}_{a}(r) \right) {} \\
\rho^{T}_{B}(x) & = & \sum_{\alpha}^{occ}\frac{2j_{a}+1}{4\pi r^{2}}2G_{a}(r)F_{a}(r) {} \\
\rho_{3}(x) & = & \sum_{\alpha}^{occ}\frac{2j_{a}+1}{4\pi r^{2}} (2t_{a}) \left( G^{2}_{a}(r)+F^{2}_{a}(r) \right) {} \\
\rho^{T}_{3}(x) & = & \sum_{\alpha}^{occ}\frac{2j_{a}+1}{4\pi r^{2}} (2t_{a})2G_{a}(r)F_{a}(r){} 
\end{eqnarray}
where the sum goes over the occupied orbitals.
In addition to these, the charge density is composed of two pieces: a direct nucleon charge density and the vector meson contribution
\begin{equation}
\label{eq:charge dens}
\rho_{\mathrm{chg}}\equiv \rho_{d}(\mathbf{x})+\rho_{m}(\mathbf{x})
\end{equation}

The direct part is given by
\begin{eqnarray}
\rho_{d}(x) &=& \rho_{p}(x)+ \frac{1}{2M} \mathbf{\vec{\nabla} \cdot} (\rho^{T}_{\mathrm{a}}(x)\mathbf{\hat{r}}) \nonumber \\
& &  {}+\frac{1}{2M^{2}}[\beta_{s}\nabla^{2}\rho_{B}(x)+\beta_{v}\nabla^{2}\rho_{3}(x)]{}
\end{eqnarray}
and the vector meson contribution arising from the coupling of the neutral vector mesons to the photon (VMD) takes the form
\begin{equation}
\rho_{m}(x)=\frac{1}{g_{\gamma} g_{\rho}} \nabla^{2}R+\frac{1}{3g_{\gamma} g_{v}} \nabla^{2}W
\end{equation}

Here the point proton density $\rho_{p}$ and nucleon tensor density $\rho^{T}_{\mathrm{a}}$ are given by
\begin{equation}
\rho_{p}=\frac{1}{2}(\rho_{B}+\rho_{3})
\end{equation}
\begin{equation}
\rho^{T}_{\mathrm{a}}=\sum^{occ}_{\sigma} \psi^{\dagger}_{\sigma}(x)i\lambda \beta \mathbf{\vec{\alpha} \cdot \hat{r}} \psi_{\sigma}(x)
\end{equation}

To solve the meson equations a Green's function method is used. The equations are iterated until a consistent solution for all the meson fields is obtained, that is, we obtain a $0^{\textit{th}}$-order solution to the meson equations by neglecting all non-linear terms on the r.h.s., and we then use this to obtain a $1^{\textit{st}}$-order solution, continuing until convergence is achieved. Because NDA guarantees that each additional term on the rhs of Eqs. (\ref{eq:Phi-equation}) - (\ref{eq:A-equation}) is smaller than the previous one, convergence is both expected and obtained.

The whole system of Eqs. (\ref{eq:G-equation}), (\ref{eq:F-equation}) and (\ref{eq:Phi-equation}) - (\ref{eq:A-equation}) is solved self-consistently until a global convergence is reached.

The values of the various constants used in this paper are given in Table~\ref{tab:parms} with a label indicating the parameter set. These values are taken from~\cite{CEL,Ser-Wal}. The constants were obtained by fitting several nuclear properties~\cite{CEL}. From the 16 constants in \ref{tab:parms} the last 3 where determined by fitting to electromagnetic properites of the nucleon.  The sets labeled W1, Q1, Q2, G1 and G2 were obtained using a weighted generalized $\chi^{2}$ defined by 
\[
\chi^{2} = \sum_{i} \sum_{X} [\frac{X^{(i)}_{\mathrm{exp}} - X^{(i)}_{\mathrm{th}}}{W^{(i)}_{X}X^{(i)}_{\mathrm{exp}}}]^{2}
\]
to fit a total of 29 observables listed as follows:

\begin{itemize}
\item The binding energies per nucleon E/B with W=0.15\%;
\item The rms charge radii $\langle r^{2}\rangle_{\mathrm{chg}}^{\frac{1}{2}} $ with W=0.2\%;
\item the diffraction-minimum-sharp, d.m.s., radii R$_{\mathrm{dms}}$ with W=0.15\%;
\item The spin-orbit splittings $\Delta E_{SO}$ of the least-bound proton and neutrons with W=5\% for $^{16}\textrm{O}$, 15\% for $^{208}\textrm{Pb}$, 25\% for $^{40}\textrm{Ca}$ and $^{48}\textrm{Ca}$, and 50\% for $^{88}\textrm{Sr}$; \footnote{Note that the spin-orbit splitting arises predominantly from the mean isoscalar scalar and vector fields that are determined by the bulk properties~\cite{Wal-book,Ser-Wal}.}
\item The proton energy $E_{p}(1h_{9/2})$ and the proton level splitting $E_{p}(2d_{3/2})-E_{p}(1h_{11/2})$ in $^{208}\textrm{Pb}$ with W=5\% and 25\%, respectively;
\item The surface-energy and symmetry-energy deviation coefficients $\delta a_{2}$ and $\delta a_{4}$ with a weight W=0.08.
\end{itemize}

The parameter set L2 was obtained by fitting the constants to reproduce the infinite nuclear matter properties~\cite{Ser-Wal}:
\begin{itemize}
\item Saturation density $\rho^{0}_{B}=0.1484\ \textrm{fm}^{-3}$.
\item Binding energy per nucleon E/B = -15.75 MeV.
\item Bulk symmetry energy = 35 MeV.
\item rms charge radius of $^{40}Ca$ = 3.482 fm.
\end{itemize}

%

\begin{table*}
\caption{\label{tab:parms} Values of the different parameters in each parameter set used to solve the self-consistent mean-field equations from~\cite{CEL,Ser-Wal}. $m_{s}$ is in MeV.}
\begin{ruledtabular}
\begin{tabular}{|c|ddddddd|}
Constant & L2 & NLC & W1 & Q1 & Q2 & G1 & G2 \\
\hline
$m_{s}$ & 520 	& 500.8  & 566.26& 504.6 & 509.6 & 506.7& 520.3 \\
$g_{s}^{2}$ & 109.63& 95.11  &138.93  & 103.6686 & 97.7699 & 97.39 &110.16 \\
$g_{v}^{2}$ & 190.43& 148.93 & 203.97 & 164.6963 & 149.2 & 147.09& 162.88\\
$g_{\rho}^{2}$ & 65.23 & 74.99  & 95.551 & 77.9558 & 73.2256 & 77.033& 89.936\\
$\eta_{1}$ &  	& 	 &  	  & 	& & 0.07060& 0.64992\\
$\eta_{2}$ &  	& 	 &  	  & 	& & -0.96161& 0.10975\\
$\kappa_{3}$ &  	& 1.9194 &  	  & 1.6582 & 1.7424 & 2.2067& 3.2467\\
$\kappa_{4}$ &  	& -7.3923&  & -6.6045 & -8.4836& -10.090& 0.63152\\
$\zeta_{0}$ &  	& 	 &  	  & 	& -1.7750 & 3.5249& 2.6416\\
$\eta_{\rho}$ &  	& 	 &  	  & & & -0.2722& 0.3901\\
$\alpha_{1}$ &  	& 	 &  	  & & & 1.8549& 1.7234\\
$\alpha_{2}$ &  	& 	 &  	  & & & 1.788& -1.5798\\
$f_{v}$ &  	& 	 &  	  & & & 0.4316& 0.6936\\
\hline
$f_{\rho}$ &  	& 	 & 3.7328 & 4.1328 & 4.2640 & 4.1572& 3.8476\\
$\beta_{s}$ &  	& 	 & -0.38482 & -0.10689 & 0.01181& 0.02844& -0.09328\\
$\beta_{v}$ &  	& 	& -0.54618 & -0.26545 & -0.18470& -0.24992& -0.45964\\
\end{tabular}
\end{ruledtabular}
\end{table*}

\section{Results}
\begin{figure}
\includegraphics[angle=-90,width=0.5\textwidth]{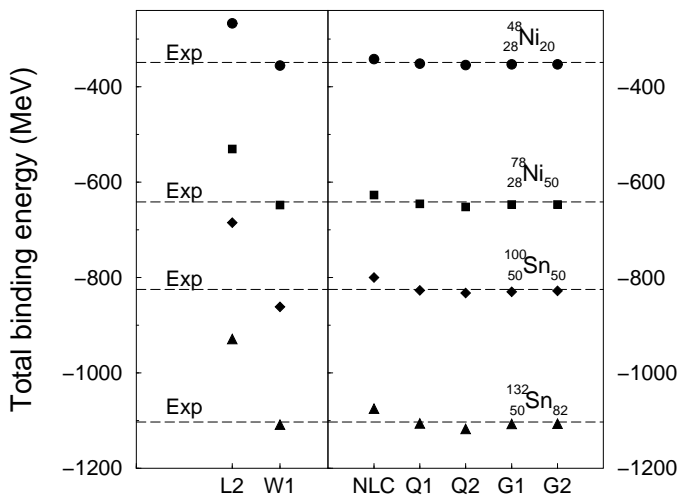}
\caption{Total binding energy of the selected nuclei. The left hand side shows two models with the same level of approximation but fit to different data sets (see text). The right hand side shows models with increasing level of approximation. NLC and Q1 have the same level of approximation as do G1 and G2.}
\label{fig:total-be}
\end{figure}

In this section we show the results of the calculations of several ground-state properties of the nuclei $^{132}_{\ 50}\textrm{Sn}_{82}$, $^{100}_{\ 50}\textrm{Sn}_{50}$, $^{48}_{28}\textrm{Ni}_{20}$, and $^{78}_{28}\textrm{Ni}_{50}$ and of the neighboring nuclei differing by one particle or one hole. We start by considering results for the total binding energy, which are shown in Fig.~\ref{fig:total-be}

 The plot is divided in two parts which show different forms of the convergence of the models. On the left are the results of total binding energy using the simplest lagrangian, i.e. with no non-linear terms. The two parameter sets L2 and W1 represent the result of fitting the constants of the model to different nuclear properties. As described in the previous section, the W1 set is fit to more data than L2. The improvement of the results is clear. The experimental result is shown as a long-dash line in the plot. The error bars are of the same size as the thickness of the line showing the experimental value. The experimental results are taken from~\cite{Web}, except for $^{48}_{28}\textrm{Ni}_{20}$ where we use the value quoted in~\cite{Brown}. On the right side of Fig. \ref{fig:total-be} are the results obtained from the more elaborate lagrangians, which in general include different non-linear interactions as shown in the previous section (see Eqns. (\ref{eq:Phi-equation}) - (\ref{eq:A-equation})). The first two points, corresponding to the NLC and Q1 parameter sets, represent the same level of approximation of the lagrangian. We see here, again, that the use of an extended set of observables in the fitting procedure dramatically improves the results. In this case, the Q1 set already gives an excellent result for all nuclei.

\begin{figure}[h]
\includegraphics[angle=-90,width=0.45\textwidth]{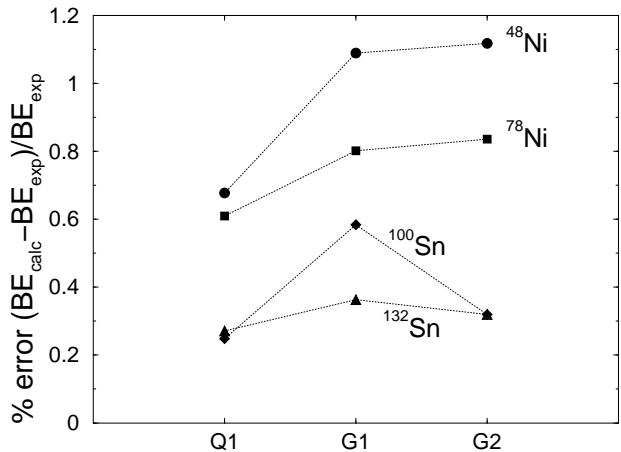}
\caption{\label{fig:percent-doubly}Percentage deviation of total binding energy of the four selected doubly-magic nuclei. Parameter sets Q1, G1 and G2 are shown.}
\end{figure}

As we move to the right on this part of the plot, we see the effect of additional interaction terms in the lagrangian. The inclusion of a fourth-order vector field term in the lagrangian, see Eq.(\ref{eq:W-equation}) and Table~\ref{tab:parms}, gives a larger binding energy. This effect is further balanced by additional non-linear terms in the G1 and G2 sets.  At this level, the sets G1 and G2 give almost the same results and cannot be differentiated. Since both sets correspond to the same level of approximation, we are looking here at the sensitivity of the results to the variations of the fitted constants.
The same features are verified for all the nuclei tested. Figure~\ref{fig:percent-doubly} shows the percentage deviation in estimating the total binding energy of the selected nuclei. We observe that it lies below approximately 1\% in all cases. In all cases the parameter set Q1 gives the smallest deviation. Extending the parameter set to G1 and G2, while validating the NDA and the assumption of ''naturalness'' as discussed in [1], does not significantly improve the quality of the fit to this quantity for the nuclei shown. For the four doubly-magic nuclei tested we obtain better agreement for the case of the Sn-isotopes than for those of Ni.

\begin{figure}[h]
\includegraphics[angle=-90,width=0.45\textwidth]{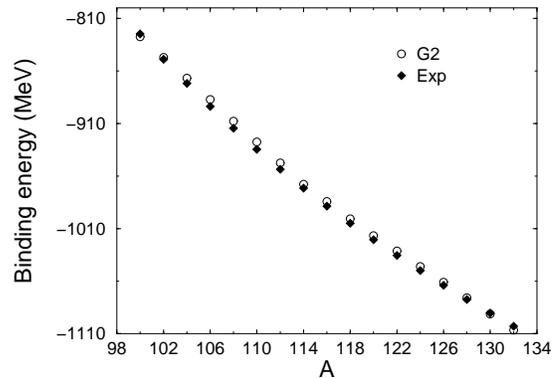}
\caption{\label{fig:be-sn-even-even}Comparison between experimental and calculated total binding energies for Sn-isotopes using the G2 parameter set.}
\end{figure}

\begin{figure}[h]
\includegraphics[angle=-90,width=0.45\textwidth]{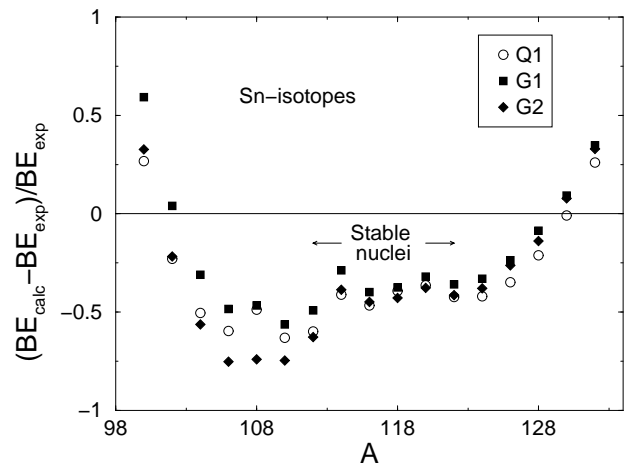}
\caption{\label{fig:percent-even-even}Percentage deviation of the total binding energy for Sn-isotopes using Q1, G1 and G2 parameter sets. The stable isotopes are indicated in the plot.}
\end{figure}

For the Sn isotopes, we extended the calculations of the total binding energy to cover the entire range of even-even nuclei. Figure~\ref{fig:be-sn-even-even} shows the results for the G2 parameter set for the isotopes $^{100}_{\ 50}\textrm{Sn}_{50}$ - $^{132}_{\ 50}\textrm{Sn}_{82}$. 
The percentage deviation in this extended case is shown in Figure~\ref{fig:percent-even-even}. Here we observe that the deviation for the three sets Q1, G1 and G2 is in absolute value less than 1\%. In general, all three parameter sets give less binding energy than observed. For all parameter sets shown in this figure, G1 gives overall the smallest deviation, except for the two nuclei $^{100}_{\ 50}\textrm{Sn}_{50}$ and $^{132}_{\ 50}\textrm{Sn}_{82}$, which lie at the extreme edges of the plot. This would imply that there is indeed improved convergence overall in going from Q1 to the extended parameter sets G1 and G2, and G1 gives better results than G2 over the entire range.

From this figure we can observe the effects of the accuracy of the isovector contributions to the effective lagrangian. Following, for example, the G1 results we observe that in the center of the plot there is essentially a constant deviation from the experimental values, while this increases close to the edges of the isotope range. A  study of the problem of the isovector contribution to the description of finite nuclei has been presented recently by Furnstahl~\cite{Furnstahl-isovector}.

We can conclude that a more sophisticated lagrangian, fit to stable nuclear properties, can reproduce ground-state total binding energies of nuclei far from stability.

It is important to note that even though the constants of the model were fit using only \textit{stable} doubly-magic nuclei, the results for binding energies of these nuclei far from stability are remarkably good.

\begin{figure}
\includegraphics[angle=0,width=0.45\textwidth,height=0.24\textheight]{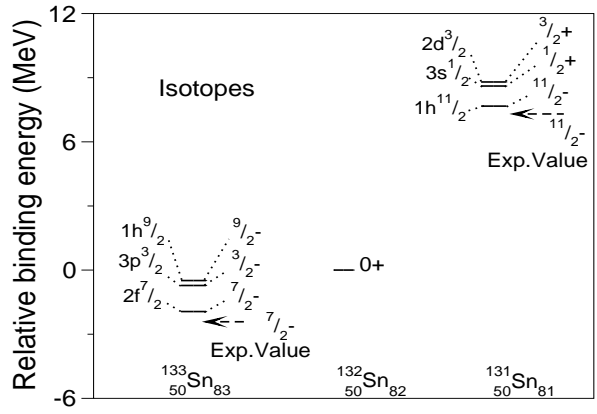}%
\caption{Level spectrum of isotopes of $^{132}_{\ 50}\textrm{Sn}_{82}$ differing by one neutron.}
\label{fig:sn132-isotopes}
\end{figure}

\begin{figure}
\includegraphics[angle=0,width=0.45\textwidth,height=0.24\textheight]{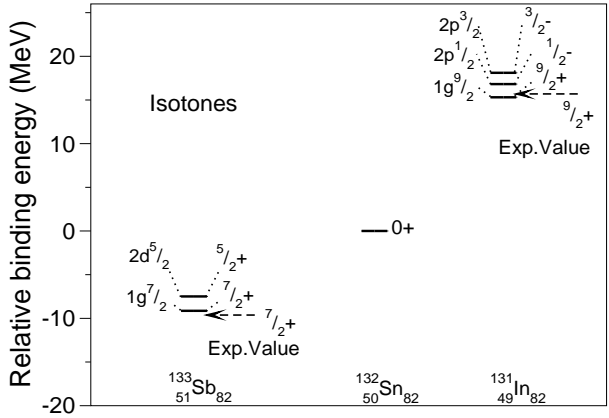}
\caption{Level spectrum of isotones of $^{132}_{\ 50}\textrm{Sn}_{82}$ differing by one proton}
\label{fig:sn132-isotones}
\end{figure}

\begin{figure}
\includegraphics[angle=0,width=0.45\textwidth,height=0.25\textheight]{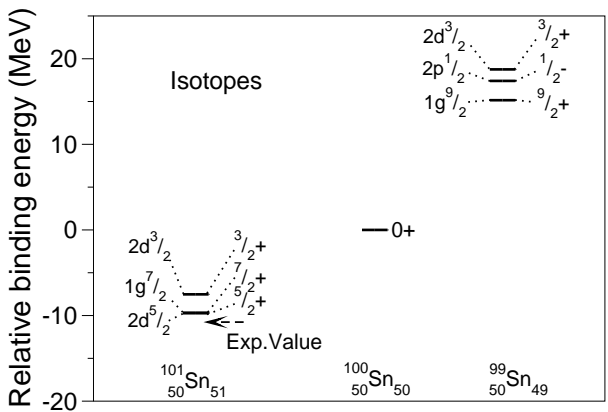}
\caption{Level spectrum of isotopes of $^{100}_{\ 50}\textrm{Sn}_{50}$ differing by one neutron.} 
\label{fig:sn50-isotopes}
\end{figure}

\begin{figure}
\includegraphics[width=0.45\textwidth]{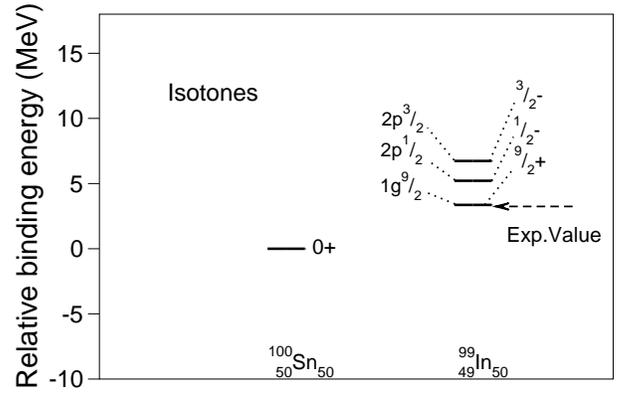}
\caption{\label{fig:sn50-isotones}Level spectrum of isotones of $^{100}_{\ 50}\textrm{Sn}_{50}$ differing by one proton.} 
\end{figure}

\begin{figure}
\includegraphics[angle=0,width=0.46\textwidth,height=0.24\textheight]{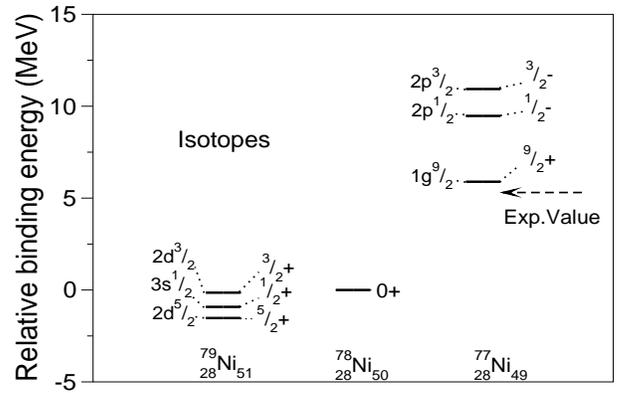}
\caption{Level spectrum of isotopes of $^{78}_{28}\textrm{Ni}_{50}$ differing by one neutron.}
\label{fig:ni78-isotopes}
\end{figure}

\begin{figure}
\includegraphics[angle=0,width=0.45\textwidth]{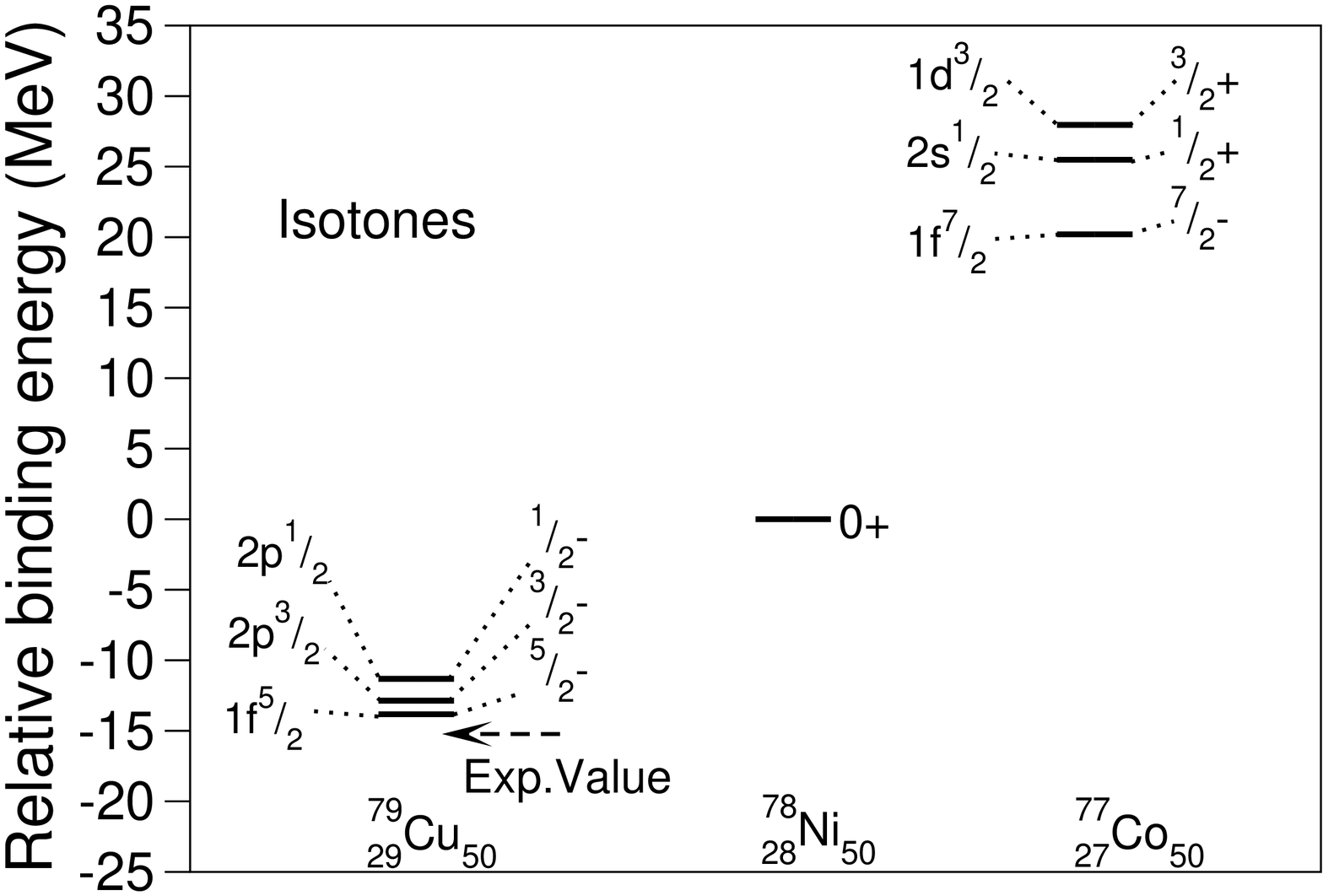}
\caption{Level spectrum of isotones of $^{78}_{28}\textrm{Ni}_{50}$ differing by one proton. }
\label{fig:ni78-isotones}
\end{figure}

Figures~\ref{fig:sn132-isotopes} through~\ref{fig:ni78-isotones} show the level structure of neighboring nuclei differing by a single particle or hole, relative to the doubly-magic nuclei. They are separated in isotopes and isotones. The core nucleus is plotted in the center of the plot marked with $0^{+}$. On the left we have put neighboring nuclei with one more particle. On the right the corresponding nucleus with one less particle is shown. When available, we include the experimental values. These results were obtained using the G2 parameter set, although the G1 set gives very similar ground state binding energies. In all these cases there is excellent agreement with the experimental data~\cite{Web}.
 


In the case of $^{131}_{\ 50}\textrm{Sn}_{81}$ the above reference gives as ground-state a value of $\frac{3}{2}^{+}$, but recent experimental results from Bhattacharyya, \textit{et al.}~\cite{Bhatt} indicate that the ground-state is in fact $\frac{11}{2}^{-}$ in accord with our results. 

We have applied the chiral effective lagrangian to calculate the ground-state properties of selected doubly-magic nuclei. The inclusion of additional interaction terms in the effective lagrangian, consistent with NDA, gives results that approach asymptotically to the experimental value for the total binding energy. We extended this calculations to neighboring nuclei, differing by one particle or one hole, and determined the binding energy relative to the doubly-magic one and found an overall agreement of better than 1\% for the total binding energy of the doubly-magic nuclei and better than 10\% for the relative binding energy of the neighboring nuclei. In this latter case, the spin and parities of the ground-states are correctly predicted. 

\begin{figure}
\includegraphics[angle=0,width=0.45\textwidth,height=0.24\textheight]{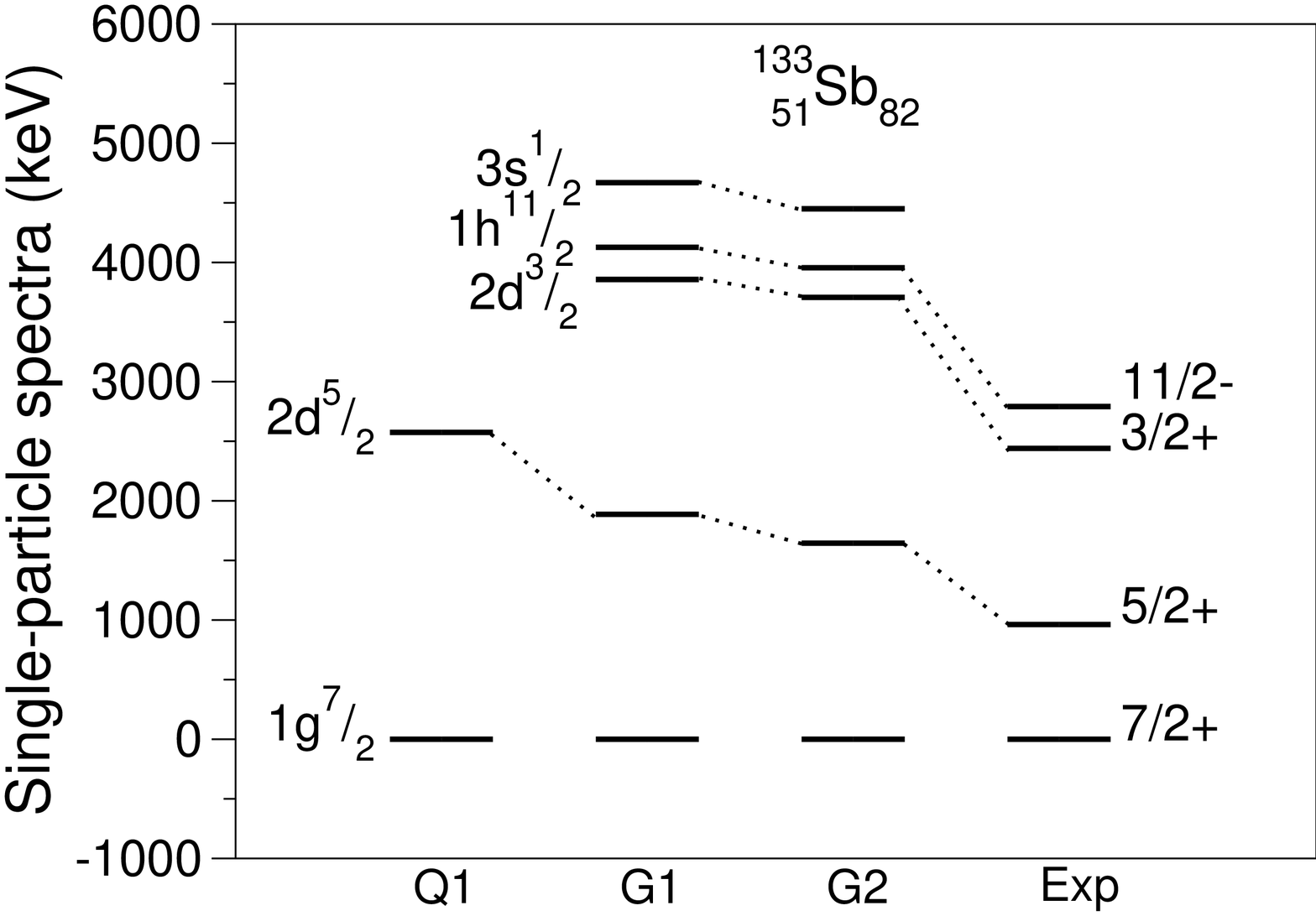}%
\caption{Comparison between single-particle level spectrum of $^{133}_{\ 51}\textrm{Sb}_{82}$ and recent experimental data}
\label{fig:sb-levels}
\end{figure}

\begin{figure}
\includegraphics[angle=-90,width=0.45\textwidth]{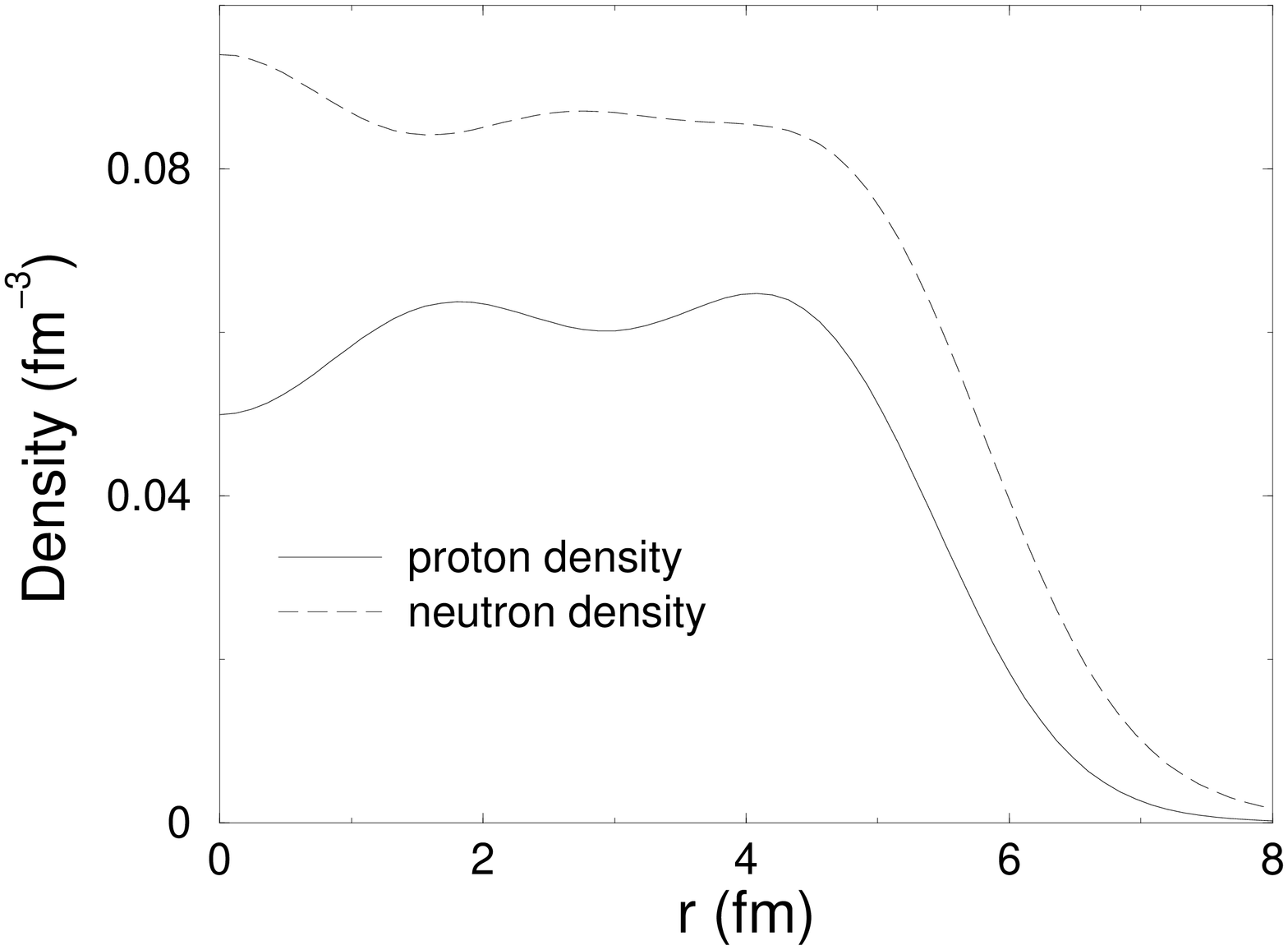}
\caption{Proton and neutron densities for $^{132}_{\ 50}\textrm{Sn}_{82}$}
\label{fig:sn132-dens}
\end{figure}

\begin{figure}
\includegraphics[angle=-90,width=0.55\textwidth]{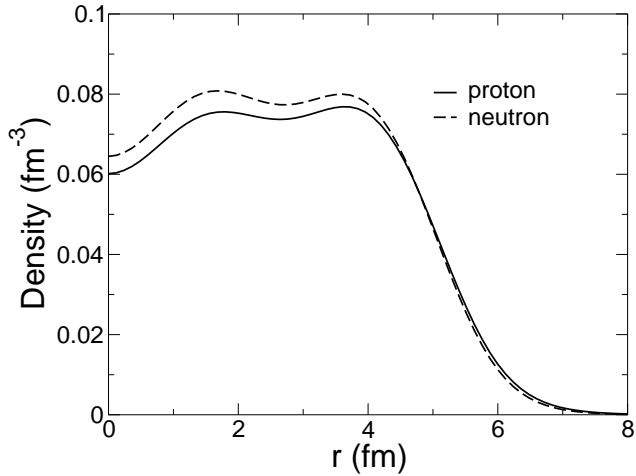}
\caption{Proton and neutron densities for $^{100}_{\ 50}\textrm{Sn}_{50}$}
\label{fig:sn100-dens}
\end{figure}

\begin{figure}[h]
\includegraphics[angle=-90,width=0.45\textwidth]{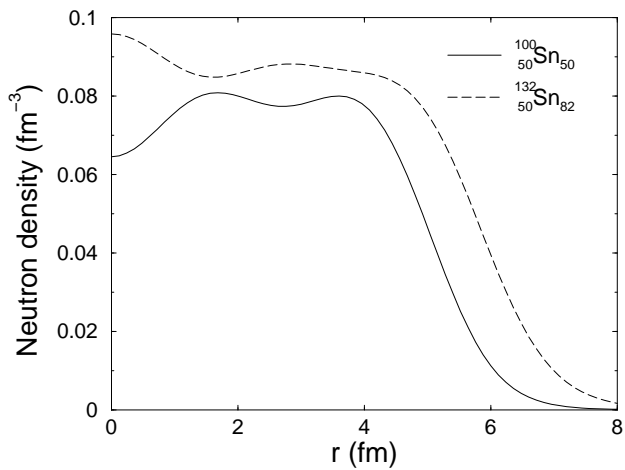}
\caption{Comparison between neutron densities of $^{132}_{\ 50}\textrm{Sn}_{82}$ and $^{100}_{\ 50}\textrm{Sn}_{50}$.}
\label{fig:sn-dens-comp}
\end{figure}

Single-particle excitation spectra for the nuclei $^{133}_{\ 50}\textrm{Sn}_{83}$ and $^{133}_{\ 51}\textrm{Sb}_{82}$ have been measured by Hoff, \textit{et al.}~\cite{Hoff} and by Sanchez-Vega \textit{et al.}~\cite{Sanchez}. The level spectra for these nuclei is shown as a function of parameter set in Figs.~\ref{fig:sb-levels} and \ref{fig:sn133-levels}. Here we chose only the G2, G1 and Q1 sets since these give the best results for the total binding energy of $^{132}_{\ 50}\textrm{Sn}_{82}$. From the discussion of the DFT approach we do not expect these single-particle and single-hole excitation spectra to be well reproduced, though we observe that there is a slow convergence of the results with higher levels of approximation. This might indicate the need for many more terms in the lagrangian. On the other hand the excellent agreement shown in Fig~\ref{fig:be-sn-even-even} for the total binding energy, and Figs. \ref{fig:sn132-isotopes} and \ref{fig:sn132-isotones} for the ground-state properties (i.e. chemical potential leading to the neighboring single-particle and single-hole nuclei) suggests that the current level of approximation is enough to reproduce the ground-state observables the theory is designed for. We observe that increasing levels of approximation give better results in the level spacing, e.g. Fig~\ref{fig:sb-levels} though the ordering is essentially independent of the level of approximation. Note that in the case of $^{133}_{\ 51}\textrm{Sb}_{82}$ we do not include results for the Q1 parameter set for levels above 2d5/2. This is because the program used did not find those states to be bound, a problem that is not present in the case of $^{133}_{\ 50}\textrm{Sn}_{83}$. This feature is expected since the higher order terms included in the G1 and G2 parameter sets  do not only contribute to the bulk properties of the nuclei but also improve the description of the single-particle spectra.
\begin{figure}
\includegraphics[angle=0,width=0.45\textwidth,height=0.24\textheight]{plot-sn133-comp-q1g1g2.eps}
\caption{Comparison between single-particle level spectrum of $^{133}_{\ 50}\textrm{Sn}_{83}$ with recent experimental data}
\label{fig:sn133-levels}
\end{figure}
In Figs.~\ref{fig:sn132-dens} and~\ref{fig:sn100-dens} we show the results for the neutron and proton densities of $^{132}_{\ 50}\textrm{Sn}_{82}$ and  $^{100}_{\ 50}\textrm{Sn}_{50}$. In both  cases the density of neutrons is always higher in the interior of the nuclei. This is particularly interesting in the case of $^{100}_{\ 50}\textrm{Sn}_{50}$ since there are an equal number of neutrons and protons implying that the protons are pushed out, due to the Coulomb potential, forming a thin layer at the surface. In contrast, in the neutron rich case the neutrons extend far out and form a neutron layer. In Fig.~\ref{fig:sn-dens-comp} we compare the neutron densities of these two doubly-magic Sn nuclei.  These results are in good agreement with calculations made using other approaches~\cite{Lalazissis}.

The calculations of the nuclear properties using the DFT approach do not include the contributions of pairing interactions. Even though most modern calculations use some type of pairing, as mentioned before, the goal of this study is to determine the convergence of the results for the optimal observables as new interaction terms are included. Extending these calculations beyond these observables gives less satisfactory results, as has been seen in the single-particle and single-hole excitation spectra of neighboring nuclei, Figs~\ref{fig:sb-levels} and \ref{fig:sn133-levels}. Although beyond the original purpose of the paper, it is interesting to compare the experimental values of the two-neutron separation energy $S_{2N}$ for the series of nuclei $^{100}_{\ 50}\textrm{Sn}_{50}$ - $^{136}_{\ 50}\textrm{Sn}_{86}$ with our calculated results. We give results for G2 only, since G1 gives essentially the same values. To make a reasonable comparison, we extract the configuration energy contribution coming from the mean-fields by eliminating from the experimental data the pairing contributions. To do this we use a simple solvable pairing model~\cite{Fetter-Walecka} of j-dependent strength $G_{j}$ corresponding to a pure pairing force. The result of the model is that $S_{2N}$ is related to the number of particles in the j-th sub-shell by the formula:
\[
S_{2N}=2\varepsilon_{\textrm{eff}}(j)+G_{j}(2j+5-2n)
\]
where n is the number of particles in the j-th shell. Fitting the experimental data to this formula, which indeed goes through all the experimental points, gives the values of the configuration energy $2\varepsilon_{\textrm{eff}}(j)$ and $G_{j}$ shown in Table~\ref{tab:pairing}.

\begin{figure}
\includegraphics[angle=-90,width=0.45\textwidth]{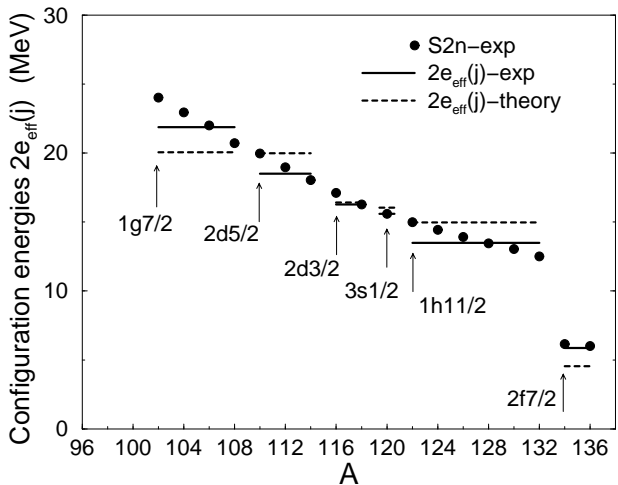}
\caption{\label{fig:conf-energy}Comparison between configuration energies for Sn-isotopes. The dashed line corresponds to calculated values using the G2 parameter set and the solid one is the value extracted from the experimental data on two-neutron separation energies $S_{2N}$ using the simple pairing force model described in the paper.}
\end{figure}

To obtain the calculated configuration energy $\varepsilon_{eff}(j)$ using the present approach, we calculated it using two methods. In one case we put only one particle in the j-th sub-shell, while the other lower shells and sub-shells where filled, and determined the total binding. Extracting the contribution from the lower filled shells and sub-shells, to the total binding energy, gave the configuration energy. In the second case the j-th sub-shell was completely filled. We proceeded as before and caclulated the total binding energy and subtracted from it the contribution from all the lower filled shells and sub-shells. This final result was averaged by dividing it by the total number of neutrons in that j-th sub-shell, $2j+1$. Both methods gave essentially the same result. Figure~\ref{fig:conf-energy} shows the comparison between the extracted and calculated configuration energies. The plot shows that, though not quantitatively, the present approach reproduces the level ordering.

\begin{table}
\caption{\label{tab:pairing} Results of extracting the configuration energy from the experimental data on $S_{2N}$ using the pairing model described in the paper.}
\begin{ruledtabular}
\begin{tabular}{|c|c|c|c|c|c|}
j & $\frac{7}{2}$ & $\frac{5}{2}$ & $\frac{3}{2}$ & $\frac{11}{2}$ & $\frac{7}{2}$ \\
$2\varepsilon_{\textrm{eff}}(j)$ & 21.88 & 18.51 & 16.27  & 13.49 & 5.86  \\
$G_{j}$ &0.2712 &0.2415 &0.2095 &0.1210 &0.03652\\
\end{tabular}
\end{ruledtabular}
\end{table}
The inclusion of pairing correlations is a subject that goes beyond the scope of the present study, but it is recognized that some sort of pairing has to be included to obtain a more complete description of nuclear observables. Work on this subject has shown that in the case of standard RMF models the approach using the DHB seems to be the most straight forward way to include such effects~\cite{Carlson}. The role of pairing in this EFT approach is currently under investigation by the author.

\section{Conclusions}
From the results shown in this paper we draw the following conclusions:
\begin{enumerate}
\item As additional interaction terms are included in the effective lagrangian, following the prescriptions given by \textit{naive dimensional analysis},  the calculated results of ground-state properties of selected nuclei far from stability converge to the observed experimental values. The overall agreement, averaged over all tested nuclei is better than 1\% in the total binding energy. The relative ground-state binding energy of neighboring nuclei, differing by one particle or one hole, are well reproduced, with a relative error of better than  10\%. The spin and parity of these nuclei, where measured, is also correctly reproduced.
\item The effect of fitting the constants of the Chiral Effective Lagrangian to a large number of observables greatly improves the results for the total binding energy of these nuclei far from stability.
\item Nucleon densities calculated with this approach are in agreement with previous works.
\item The effective field theory and density functional theory approach provides a predictive framework for dealing with QCD in the strong-coupling, nuclear physics regime~\cite{Many-body}.
\item Additional interactions to reproduce the effect of pairing are needed to extend the theory to reproduce additional ground-state properties.
\end{enumerate}

I will like to thank Dr. J. D. Walecka for his support and advice, Dr. R. J. Furnstahl for his useful insight on the subject and Dr. B. D. Serot for his comments on the paper. This work is supported in part by DOE grant DE-FG02-97ER41023.


\begin{thebibliography}{99}

\bibitem{CEL} R. J. Furnstahl, B. D. Serot, H. Tang, Nucl. Phys. A \textbf{615}, 441-482 (1997).
\bibitem{Toki} H. Toki, K. Sumiyoshi, D. Hirata, H. Shen, I. Tanihata, RIKEN Review No.26 (January, 2000): \textit{Focused on Models and Theories of the Nuclear Mass}, p. 17-22.
\bibitem{Toki-2} H. Shen, H. Toki, K. Oyamatsu, K. Sumiyoshi, Prog. Theor. Phys. \textbf{100}, 1013 (1998).
\bibitem{Furnstahl1} R. J. Furnstahl, \textit{Recent Developments in the Nuclear Many-Body Problem}, nucl-th/0109007.
\bibitem{Furnstahl} R. J. Furnstahl, B. D. Serot, Comments on Modern Physics, Vol \textbf{2}, Part A, pp.23-45.
\bibitem{Ser-Wal} B. D. Serot, J. D. Walecka,  Int. J. Mod. Phys. E \textbf{6}, 515-631 (1997).
\bibitem{Wal-book} J. D. Walecka, \textit{Theoretical Nuclear and Subnuclear Physics}, Oxford University Press (1995).
\bibitem{Furnstahl2} R. J. Furnstahl, B. D. Serot, Nucl. Phys. A \textbf{671}, 447-460 (2000).
\bibitem{Furnstahl3} R. J. Furnstahl, B. D. Serot, Nucl. Phys. A \textbf{663\&664}, 513c-514c (2000).
\bibitem{NDA-1} H. Georgi, A. Manohar, Nucl. Phys. B \textbf{234}, 189 (1984).
\bibitem{NDA-2} H. Georgi, Phys. Lett. B \textbf{298}, 187 (1993).
\bibitem{Kohn} W. Kohn, Rev. Mod. Phys. \textbf{71}, 1253-1266 (1999).
\bibitem{DFT-QHD-1} C. Speicher, R. M. Dreizler, Ann. Phys. \textbf{213}, 312-354 (1992).
\bibitem{Many-body} B. D. Serot and J. D. Walecka in \textit{150 Years of Quantum Many-Body Theory} (Edited by R. F. Bishop, \textit{et al.}, World Scientific, 2001) p. 203.
\bibitem{Lalazissis} G. A. Lalazissis, D. Vretenar, P. Ring, Phys. Rev. C \textbf{57}, 2294-2300 (1998).
\bibitem{Lenske} H. Lenske, F. Hofmann, C. M. Keil, Prog. Part. Nucl. Phys. \textbf{46}, 187-196 (2001).
\bibitem{Meng}	J. Meng, K. Sugawara-Tanabe, S. Yamaji, A. Arima, Phys. Rev. C \textbf{59}, 154-163 (1999).
\bibitem{skins} S. Mizutori, J. Dobaczewski, G. A. Lalazissis, W. Nazarewicz, P. G. Reinhard, Phys. Rev. C. \textbf{61}, 044326 (2000).
\bibitem{England} T. England, M. Hjorth-Jensen, E. Osnes, Proceedings of RNB5, April 2000, Divonne, France. To appear in Nucl. Phys. A.
\bibitem{Hofmann} F. Hofmann, C. M. Keil, H. Lenske, Phys. Rev. C \textbf{64}, 034314 (2001).
\bibitem{Grasso} M. Grasso, N. Sandulescu, Nguyen Van Giai, R. J. Liotta, Phys. Rev. C \textbf{64}, 064321 (2001).
\bibitem{Carlson} B. V. Carlson, D. Hirata, Phys. Rev. C \textbf{62}, 054310 (2000).
\bibitem{Estal}	M. Del Estal, M. Centelles, X. Vinas, S. K. Patra, Phys. Rev. C \textbf{63}, 044321 (2001).
\bibitem{Zhang}	Jing-ye Zhang, Y. Sun, M. Guidry, L. L. Riedinger, G. A. Lalazissis, Phys. Rev. C \textbf{58}, 2663-2666 (1998).
\bibitem{DFT-QHD-2} R. N. Schmid, E. Engel, R. M. Dreizler, Phys. Rev C \textbf{52}, 164-169 (1995).
\bibitem{DFT-QHD-3} R. N. Schmid, E. Engel, R. M. Dreizler, Phys. Rev C \textbf{52}, 2804-2806 (1995).
\bibitem{Bachman} Theodore Thomas Bachman, \textit{Solutions to the Relativistic Hartree Equations for nuclei far from stability}, senior thesis, College of William and Mary, (1996), unpublished.
\bibitem{GSI} Gesellschaft fur Schwerionenforschung (GSI), \textit{Proposal for an International Accelerator Facility for Research with Heavy Ions and Antiprotons}, November 2001.
\bibitem{CCWZ} C. Callan, S. Coleman, J. Wess, and B. Zumino, Phys. Rev. \textbf{177} 2247 (1969).
\bibitem{Web} Table of the Nuclides at http://www2.bnl.gov/ton/.
\bibitem{Brown} B. Alex Brown, Phys. Rev. C \textbf{58}, 1, 220-231 (1998).
\bibitem{Bhatt} P. Bhattacharyya, \textit{et al.}, Phys. Rev. Lett. \textbf{87}, 062502 (2000).
\bibitem{Hoff} P. Hoff, \textit{et al.}, Phys. Rev. Lett. \textbf{77}, 1020 (1996).
\bibitem{Sanchez} M. Sanchez-Vega, B. Fogelberg, H. Mach, R. B. E. Taylor, A. Lindroth, Phys. Rev. Lett. \textbf{80}, 5504 (1998).

\bibitem{Fetter-Walecka} A. L. Fetter, J. D. Walecka, \textit{Quantum theory of many-body systems}, McGraw-Hill (1971).
\bibitem{Furnstahl-isovector} R. J. Furnstahl, \textit{Neutron radii in mean-field models}, nucl-th/0112085.
\bibitem{deformed} R. J. Furnstahl, C. E. Price, G. E. Walker, Phys. Rev. C \textbf{36} 2590 (1987).
\end{thebibliography}

\end{document}